\documentclass[prb,aps,amsmath,amssymb,twocolumn,groupedaddress,floats,showpacs,final,reprint]{revtex4-1}
\usepackage{amsmath}
\usepackage{graphicx}
\usepackage{amssymb}
\usepackage{amscd}
\usepackage{bm}
\usepackage{epstopdf}
\usepackage{color}
\usepackage{mathrsfs}

\makeatother

\newcommand{\be}{\begin{equation}}
\newcommand{\ee}{\end{equation}}
\newcommand{\bt} { \begin{tabular} }
\newcommand{\et}{ \end{tabular} }
\newcommand{\bc} { \begin{center} }
\newcommand{\ec}{ \end{center} }

\newcommand{\bfi}{\begin{figure} }
\newcommand{\efi}{\end{figure} }

\newcommand{\btb} { \begin{table} }
\newcommand{\etb}{ \end{table} }

\begin{document}
%
\title{The Massive Goldstone (Higgs) mode in two-dimensional ultracold atomic lattice systems}

\author {Longxiang Liu}
\affiliation{National Laboratory for Physical Sciences at Microscale and Department of Modern Physics, University of Science and Technology of China, Hefei, Anhui 230026, China}

\author {Kun Chen}
\email{chenkun@mail.ustc.edu.cn}
\affiliation{National Laboratory for Physical Sciences at Microscale and Department of Modern Physics, University of Science and Technology of China, Hefei, Anhui 230026, China}
\affiliation{Department of Physics, University of Massachusetts, Amherst, Massachusetts 01003, USA}

\author {Youjin Deng}
\email{yjdeng@ustc.edu.cn}
\affiliation{National Laboratory for Physical Sciences at Microscale and Department of Modern Physics, University of Science and Technology of China, Hefei, Anhui 230026, China}
\affiliation{Department of Physics, University of Massachusetts, Amherst, Massachusetts 01003, USA}

\author {Manuel Endres}
\affiliation{Department of Physics, Harvard University, Cambridge, MA 02138, USA}
\affiliation{Institute for Quantum Information and Matter, Department of Physics, California Institute of Technology, Pasadena, CA 91125, USA}

\author {Lode Pollet}
\affiliation{Department of Physics and Arnold Sommerfeld Center for Theoretical Physics, Ludwig-Maximilians-Universit{\"a}t M{\"u}nchen, D-80333 M{\"u}nchen, Germany}

\author {Nikolay Prokof'ev}
\affiliation{Department of Physics, University of Massachusetts, Amherst, Massachusetts 01003, USA}
\affiliation{Russian Research Center ``Kurchatov Institute", 123182 Moscow, Russia }

\date{\today }
\vspace{0.2cm}
\begin{abstract}We discuss how to reveal the massive Goldstone mode, often referred to as the Higgs amplitude mode, near the Superfluid-to-Insulator quantum critical point (QCP) in a system of two-dimensional ultracold bosonic atoms in optical lattices. The spectral function of the amplitude response is obtained by analytic continuation of the kinetic energy correlation function calculated by Monte Carlo methods. Our results enable a direct comparison with the recent experiment [M. Endres, T. Fukuhara, D. Pekker, M. Cheneau, P. Schau{\ss}, C. Gross, E. Demler, S. Kuhr, and I. Bloch, Nature {\bf 487}, 454-458 (2012)], and demonstrate a good agreement for temperature shifts induced by lattice modulation. Based on our numerical analysis, we formulate the necessary conditions  in terms of homogeneity, detuning from the QCP and temperature in order to reveal the massive Goldstone resonance peak in spectral functions experimentally. We also propose to apply a local modulation at the trap center to overcome the inhomogeneous broadening caused by the parabolic trap confinement.

\end{abstract}

\maketitle
%
%
%
%
\section{Introduction}
\label{sec:1}
%
Collective modes are important for understanding dynamic susceptibilities, which include experimentally observed spectral functions and transport properties. The situation becomes particularly intriguing in strongly coupled systems, where a simple description in terms of weakly interacting excitations is unreliable and perturbation theories fail even qualitatively. Under these conditions, the role of underlying collective modes on dynamic susceptibilities becomes increasingly more important but is hard to calculate~\cite{book}. This is exactly what happens in the vicinity of the two-dimensional (2D) Superfluid-to-Mott insulator quantum critical point (SF-MI QCP)~\cite{fisher}. Though it is considered to be one of the best studied strongly coupled systems,  its quantum critical dynamics is still poorly understood, both theoretically and experimentally.

In superfluids near SF-MI quantum criticality, the effective field theory in terms of a complex scalar order parameter $\Psi$ features an emergent particle-hole symmetry and Lorentz invariance, and is expected to have
two types of collective modes.\cite{Goldstone}  The first one originates from fluctuations of the phase of $\Psi$ and describes a massless Bogoliubov-Nambu-Goldstone mode. The second one describes amplitude fluctuations and is associated with a massive Goldstone (MG) mode\cite{Goldstone}, often referred to as a Higgs amplitude mode. The fate of the MG-mode in 2D is an intriguing and controversial issue because its decay into lower-energy gapless modes is found to be strong. The mode was argued to become either completely overdamped ({\it i.e.}, without any resonance type feature in spectral functions~\cite{old_papers}) or be detectable as
a well-defined resonance peak in certain spectral functions on the superfluid side away from (but not on approach to) the critical point~\cite{huber1, huber2, pod11}.

Recent progress on ultracold atom in optical lattice experiment ~\cite{bloch}, as well as Monte Carlo simulations~\cite{Lode, Snir1, Kun, Snir2}, $1/N$ corrections to higher order \cite{pod12}, and non-perturbative renormalization group methods~\cite{Rancon, rose} have presented solid evidence in favor of yet another scenario: A critical resonance in the universal scaling regime. Unlike MG-modes in three (and higher) dimensions which become sharper when approaching the QCP, the ratio of the width of the resonance peak over its mass remains constant in (2+1) dimensions. 

In order to detect the MG-mode experimentally, the scalar spectral function $A(\omega)$ ({\it i.e.}, the correlation function of $|\Psi^2|$) is considered to be the best probe~\cite{pod11}.
On the SF side of the transition point, in the scaling limit, $A(\omega)$ takes the form \cite{book}
\begin{equation}\label{eq:scaling}
A_{\rm SF}(\omega) \propto \Delta^{3-2/\nu}\Phi_{ \rm SF} (\frac{\omega}{\Delta}) \;,
\end{equation}
where $\Delta$ is the Mott insulator gap for the same amount of detuning from the QCP, and $\nu=0.6717$ is the correlation length exponent for the $U(1) \equiv O(2)$ universality in $(2+1)$ dimensions~\cite{Machta,Vicari}.
The universal function $\Phi_{\rm SF} (x)$ starts as $\Phi_{\rm SF} (x\to 0) \propto x^3$ and saturates to a quasi-plateau with weak $\omega$ dependence $\Phi_{\rm SF} (x \gg 1 ) \propto x^{3-2/\nu} \approx x^{0.0225}$ . The intermediate regime between the two limits can be constructed numerically, where a well-defined resonance peak associated with the critical MG-mode is observed at $x=3.3(8)$\cite{Kun}. Monte Carlo data also suggest that a
similar universal resonance, though less pronounced, may equally well be seen on the other side of the transition, as well as in the quantum critical normal liquid~\cite{Kun, Snir2, Rancon}. These conclusions are yet to be confirmed or refuted experimentally.
The bottleneck of the numerical analysis is analytical continuation of data for correlation functions from imaginary
to real frequencies, $A(i\omega_n) \to A(\omega) $, where $i\omega_n = 2\pi n T$ with $n=0, \pm 1, \pm 2 \dots$ are
Matsubara frequencies. This procedure is a notorious, ill-posed problem that requires application of certain regularization schemes \cite{maxent, nikolay}, and thus independent experimental studies are required for final understanding.

The ultracold atom experiment of Ref.~\onlinecite{bloch} aimed at detecting the MG-mode 
in $A_{\rm SF}(\omega)$
and  confirmed the expected softening of the quantum critical spectrum
implied by (\ref{eq:scaling}) but remained inconclusive with regards to the existence of a well-defined MG resonance. To obtain $A(\omega)$, a 2D Bose Hubbard system was gently "shaken" by modulating
the lattice laser intensity (lattice depth) and probed by \emph{in situ} single-site- and single-atom-resolved measurements.
The observed signal (through temperature increase) featured a broad maximum whose onset softened on approach to QCP, in line with the scaling law (\ref{eq:scaling}). The onset correlates remarkably well with the energy of the MG-mode,  while the ratio of the onset width to its frequency was measured to be approximately constant when approaching the critical point. However, a resonance-type peak 
with diminishing width was not detected, which can be interpreted
either as evidence for the MG-mode being overdamped in the critical state or as broadening caused
by finite temperature and system inhomogeneity (tight confinement) effects~\cite{Lode}. 
Thus a direct comparison between numerical calculations and experimental measurements 
with a common setup is crucial to settle the controversy.

In this paper, we employ an \textit{ab initio} numerical procedure based on quantum Monte Carlo simulations and numerical analytic continuation~\cite{nikolay} to calculate spectral functions for ultracold atoms in optical lattices.
The final result for the temperature increase as a function of modulation frequency successfully reproduces the
main data of Ref. \onlinecite{bloch} for the experimental setup ``as is''; {\it i.e}, in the spirit of the quantum simulation paradigm~\cite{lode12}. The consistency between numerical results and experimental measurements establishes the reliability of both approaches, and, in particular, validates the analytic continuation procedure.
Moreover, simulations performed for various system parameters indicate several improvements/requirements with regards to the experimental setup that will help revealing the resonance peak in the spectral function.
They include (i) the system should be effectively homogeneous to avoid inhomogeneous broadening, which can be achieved through confining the lattice depth modulation locally at the parabolic trap center; (ii) the detuning from the QCP should be small, $j/j_c \le 1.05$, where $j=J/U$ is the dimensionless coupling parameter
for the Bose Hubbard model introduced below [see Eq.(\ref{eq:BH})], 
and $j_c$ is its critical value; and (iii) the system's temperature should be at least as low as the Berezinskii\text{-}Kosterlitz\text{-}Thouless transition point $T_c$. 
Our results suggest that a direct observation of a well-defined resonance peak and understanding the fate of the MG-mode experimentally is challenging but not impossible.

The rest of the paper is organized as follows. We introduce the model in Sec.\ref{sec:model} and describe the  numerical procedure in Sec.~\ref{sec:theory}. The comparison between the temperature response from simulations and experimental measurements in a specific setup from Ref. \onlinecite{bloch} is presented in Sec.\ref{sec:simulation}. We discuss requirements and possible experimental improvements to reveal the MG resonance in  Sec.\ref{sec:observation}.

\section{the model}
\label{sec:model}
%

Ultracold bosons in optical lattices offer unique possibilities to study the SF-MI quantum phase transition in 2D~\cite{spielman, bloch_review}.  At low enough temperatures the physics of the system is restricted to the lowest Bloch band, and can be described by the Bose-Hubbard (BH) model ~\cite{fisher, zoller}, which is parametrized by a hopping amplitude $J$ and an on-site interaction energy $U$,
\be
\label{eq:BH}
\hat{H}_0 = - J \sum_{\left<i,j \right>}  (b_i^\dag b_j^{\,}+h.c.)+\frac{U}{2}\sum_{i} n_i(n_i-1)-\sum_{i}(\mu- V_i) n_i ,
\ee
where $b_i^\dag$ ($b_i$) creates (annihilates) a particle on the site  $i$, and $\left< i, j \right>$ denotes the sum over nearest neighbors on the square lattice. In the BH model, the dimensionless coupling parameter $j = J/U$ is easily tunable via the lattice depth, and the dimensionality of the system can be reduced to 2D by suppressing the hopping in the third direction. The total particle number $N$ is controlled by the global chemical potential $\mu$. Finally, an ultracold atomic gas is trapped by a confining potential $V_i$, which is usually harmonic, $V_i = \frac{1}{2}m \omega^2 d^2 R_i^2$ (with $m$ the mass of atom, $d$ the lattice spacing, and $R_i$ the distance of site $i$ from the trap center measured in units of $d$). Within the local density approximation picture (LDA), 
$\mu-V_i$ plays the role of a local chemical potential. 

Ideally, without the $V_i$ term in Eq. (\ref{eq:BH}), the system is a homogeneous 2D Bose Hubbard model and its phase diagram is known with high accuracy \cite{QCP1, QCP2, QCP3} at both zero and finite temperature. In the ground state, the system undergoes a second order phase transition from the SF to the MI when decreasing the ratio $j=J/U$ at fixed integer filling factor. At filling factor $\langle n\rangle =1$, the transition occurs at $j_c^{-1}=16.7424(1)$ and $\mu_c/J=6.21(2)$ and features a QCP with emergent particle-hole symmetry, which enlarges the Galilean invariance to Lorentz invariance (the system is actually conformally invariant). The SF phase is supposed to have
the critical MG-mode according to the discussion in the previous section.

 When the $V_i$ term is presented  as in current experimental implementations~\cite{bloch}, due to the inhomogeneous local chemical potential, the particle density decreases to zero when moving away from the center of the trap. Any conclusion regarding the existence of the GM-resonance in the homogeneous case cannot be naively applied to the realistic experiment, even if the center
of the atomic system is fine-tuned to be in the vicinity of QCP. A careful bottom-up calculation of the scalar spectral function is required in order to understand the experimental signal.
\section{Spectral function measurement: Theory\label{sec:theory}}
In this section, we revisit the generic mathematical framework for the measurement of the scalar spectral function in ultracold atoms. \\

In the BH model, the total kinetic energy  $\hat{K}=- J \sum_{\left< i,j \right>}  (b_i^\dag b_j^{\,}+h.c.)$ is the simplest operator with nontrivial dynamics leading to strong scalar response. Thus, we may consider adding an external perturbation term $\delta f(t) \hat{K}$ to the Hamiltonian (\ref{eq:BH}). Within standard linear response theory, the total kinetic energy response is proportional to the external field, and the ratio defines the response function $\chi(\omega, T) \equiv \delta \langle \hat{K}(\omega)\rangle_T/\delta f(\omega)$ where $\langle ... \rangle_T$ denotes the thermal average at temperature $T$. The spectral function is defined as the dissipative part of the response function, $A(\omega, T) \equiv 2 {\rm Im} \chi(\omega, T)$, so that $A(\omega, T)$ is proportional to the energy absorbed by the system, which, in turn, determines the temperature change of the system. To learn about the spectral function, one can measure either the total kinetic energy response or the temperature change. Though being rather indirect, the latter one is the quantity that is measurable in the ultracold atom experiment~\cite{bloch}.

Experimentally, a small uniform modulation $\delta V_0 \cos(\omega t)$ of the optical lattice depth $V_0$ is applied in the 2D plane to generate the external perturbation term~\cite{modulation, bloch}.
In the parameter regime where the BH model is a valid approximation, the lattice depth in units of the recoil energy $E_r=\pi^2/2md^2$ is much larger than unity and controls both parameters $J$ and $U$: $J \simeq \frac{4}{\sqrt{\pi}}E_r\left( V_0/E_r \right)^{3/4}e^{-2\sqrt{V_0/E_r}} $, $U \propto (V_0/E_r)^{D/4}$ ~\cite{bloch_review} where the effective dimension $D=2$. Substituting $V_0$ with $V_0+\delta V_0 \cos(\omega t)$ in $J$ and $U$, and keeping terms to first order in $\delta V_0$, the perturbed BH Hamiltonian reads
\begin{equation}\label{eq:pBH}
\hat{H}(t) =\hat{H_0} +\delta g(t)\hat{H_0} +\delta g(t)\sum_{i} V_i\hat{n}_i +\delta f(t)\hat{K}.
\end{equation}
Here, the generalized forces $\delta f(t)=\delta f_0 \cos(\omega t)$ with $\delta f_0=(\frac{1}{4}-\sqrt{V_0/E_r})\frac{\delta V_0}{V_0}$ and $\delta g(t)=\delta g_0 \cos(\omega t)$ with $\delta g_0=\frac{1}{2}\frac{\delta V_0}{V_0}$ are linear in $\delta V_0$. Note that the second term in Eq.(\ref{eq:pBH}) commutes with $H_0$ and yields no contribution to the spectral function. Furthermore, we argue that the effect from the confining potential term (third term) is also negligible compared to the kinetic energy term (fourth term) if one of the following conditions is satisfied,  i) $\vert \delta g_0/\delta f_0 \vert =2/\vert (4\sqrt{V_0/E_r}-1) \vert \ll 1$; ii) the
trap is large enough and LDA is valid; namely, it is possible to decompose the system 
into independent mesoscopic regions whose sizes are larger than the correlation length but are still small enough to be regarded as homogeneous systems
with the local chemical potential $\mu-V_i$. This implies, on the one hand, that the total response of the system can be approximated by a sum over independent contributions from mesoscopic regions, while, on the other hand, the confining potential term in each mesoscopic region is proportional to the particle number in this region and thus commutes with the local $H_0$ (i.e., it is not dynamic under LDA). The combined effect is that the confining potential does not contribute to the linear response. The validity of LDA for critical systems has been addressed before in Ref.~\onlinecite{Pollet2010}

For ultracold atoms whose dynamics is dominated by the kinetic energy term,
the energy dissipation rate is proportional to the kinetic energy spectral function $A(\omega)$,
\begin{equation}\label{eq:Energy}
\dot{E}(\omega,T)=\frac{\omega}{4} A(\omega, T)\delta f_0^2+P(\omega,T).
\end{equation}
Here $P(\omega,T)$ is the heating power from other mechanisms (ultracold atoms are always coupled to the photon subsystem and are subject to collisions with the background gas).
In the leading approximation, $P$ does not depend on the small lattice-depth modulation, 
and we expect $P(\omega,T) \simeq P(T)$.

Assuming that the system is quasistatic, {\it i.e.}, the relaxation time is small enough, $\tau \ll 1/\omega$, the thermodynamics can applied at all times and the final temperature shift can be deduced from the following self-consistent equation,
\begin{equation}\label{eq:final}
T(\omega,t)-T(\omega, 0)=\int_0^t \frac{\dot{E}(\omega,T(\omega, t'))}{C(T(\omega, t'))} dt'
\end{equation}
where $t$ is the period of the modulation and $C(T)$
is the heat capacity. We do {\it not} assume here that under the linear response conditions
one is allowed to neglect time dependence of temperature on the r.h.s. of Eq.~(\ref{eq:final}).
This is because over long modulation times the temperature change might be substantial.   
In the experiment~\cite{bloch}, the initial temperature is chosen to be frequency independent, $T(\omega, 0)=T_{\rm ini}$, 
and the modulation time to be a certain fixed number of modulation cycles 
$t=t(\omega)=2\pi M/\omega$. 
Then the final temperature dependence on frequency,
$T_{\rm fin}(\omega)=T(\omega, t(\omega))$, is directly related to  
$A(\omega)$, and any sharp resonance structure in $T_{\rm fin}(\omega)$ can be traced back 
to the spectral function; i.e., the temperature response provides a practical probe to detect the MG-mode as demonstrated in Ref. \onlinecite{bloch}.

\section{Measuring the spectral function in simulations\label{sec:simulation}}
In this section, the experimental setup from Ref. \onlinecite{bloch} is used as a benchmark 
system for calculation of the temperature response from first principles.
The parameters closest to the QCP include the lattice depth $V_0=10E_r$, which gives a dimensionless coupling parameter $j/j_c=1.2$ (or $U=14J$), and the particle number $\left<N \right>=190(36)$. Combined with the unity filling requirement at the trap center, this corresponds to the harmonic confinement $V_i/J=0.0915(x_i^2 +y_i^2)$. ~\cite{Lode} The small dimensionless modulation
of the lattice depth is $\delta V_0/V_0 \simeq 0.03$, which corresponds to the generalized 
forces $\vert \delta f_0 \vert=0.087$ and $\vert \delta g_0 \vert=0.015$.  Those parameters define the perturbed BH model (\ref{eq:pBH}).

We argue that the external potential term in Eq. (\ref{eq:pBH}) is negligible, since both conditions discussed in the previous section are fulfilled, i) $\vert \delta g_0/\delta f_0 \vert =0.17 \ll 1$ ; ii) the correlation length near the trap center at a typical experimental temperature is about one lattice spacing~\cite{Lode}, so that the LDA also holds in the vicinity of the trap center, which dominates the total response.

In the experiment, the modulation protocol consists of two stages. First, ultracold atoms are modulated for $M=20$ oscillation cycles. Second, the system is held to thermalize for some time such that the sum of modulation and hold time is constant at 200 milliseconds for all modulation frequencies. During the first stage, both the modulation and the heating power $P(T)$ contribute to the system's energy dissipation, while during the second stage, only the heating power $P(T)$ contributes to the energy increase.  The integral over time in Eq. (\ref{eq:final}) must hence be divided into two segments in order to correctly account for both modulation and holding stages. The advantage of keeping the two-stage time at the same value is that the contribution of $P(T)$ is essentially constant for all modulation frequencies. Finally, the temperature $T_{\rm fin}(\omega)$  is determined by slowly ramping the system to the atomic limit and measuring the atomic parity in-situ with single-site resolution.

We rely on path-integral quantum Monte Carlo (MC) simulations with worm-type updates ~\cite{worm} to calculate the scalar spectral functions and the specific heat. Since the particle loss in the experiment is negligible, we simulate the system in the canonical ensemble.

\begin{figure}[ht]
\includegraphics[clip,angle=0,width=8.3cm]{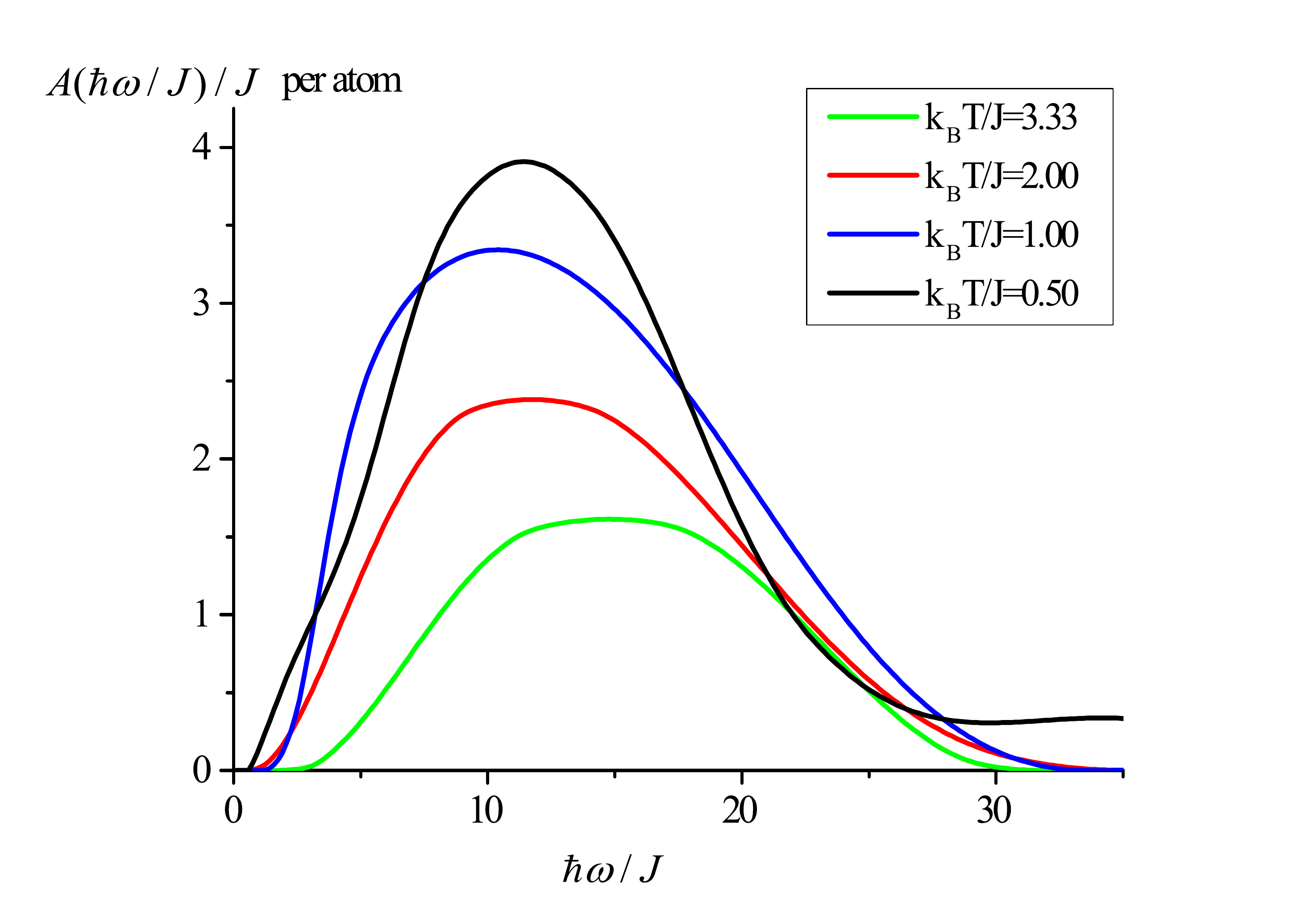}
\caption{\label{sf-diag} Spectral functions of the kinetic energy per atom at different temperatures for the experimental setup of Ref.~\onlinecite{bloch}. 
All curves look qualitatively similar. }
\end{figure}

In MC simulations, it is straightforward to measure the imaginary time correlation function for the kinetic energy, $\chi(\tau)=\langle K(\tau)K(0) \rangle_T-\langle K \rangle_T^2$, which is related to $A(\omega )$ via the spectral integral
\begin{equation}\label{ancon}
\chi(\tau )=\int_{0}^{+\infty} \frac{d\omega}{2\pi} \mathscr{N}(\tau,\omega;T) A(\omega ),
\end{equation}
with the finite-temperature kernel, $\mathscr{N}(\tau,\omega;T)=2(e^{-\omega \tau}+e^{-\omega(1/k_{\rm B} T-\tau)})/(1-e^{-\omega/k_{\rm B} T})$.
We employ the same protocol for collecting and analyzing data as in Ref.~\onlinecite{Lode, Kun}. More precisely, we collect statistics for the correlation function at Matsubara frequencies $\omega_n$
\begin{equation}\label{chi}
\chi(i\omega_n )=\frac{1}{\beta} \langle | K(i\omega_n)|^2\rangle_T+\langle K \rangle_T,
\end{equation}
and recover $\chi(\tau )$ by a Fourier transform.
In the path-integral representation, $\chi(i\omega_n )$ has a direct unbiased estimator, $\vert \sum_k e^{i\omega_n \tau_k} \vert^2$,
where the sum runs over all hopping transitions in a given configuration. 
Once $\chi(\tau )$ is obtained from $\chi(i\omega_n )$ with an accuracy up to $10^{-4}$ the analytic continuation method described in Ref.~\onlinecite{nikolay} is applied to extract the spectral function $A(\omega )$. We present the analytically continued results at different temperatures in Fig.\ref{sf-diag}, where we see that the curves look qualitatively similar at all temperatures in the range between $0.5J/k_{\rm B}$ and $3.33J/k_{\rm B}$; values of $A(\omega; T)$ at any temperature in this range can be estimated using linear interpolation. All spectral functions vanishing at zero frequency reflects the absence of dissipation in response to a static external field. Another way to understand this result is through the dissipation-fluctuation theorem, $A(\omega; T)=(1-e^{-\omega/k_{\rm B} T})S(\omega; T)$ where $S(\omega; T)$ is the dynamic correlation function of kinetic energy. Zero value of $A(\omega=0; T)$ is a natural outcome of a finite $S(\omega=0; T)$, see Fig. 6 in Ref. ~\onlinecite{Lode}. We also would like to point out that the analytical continuation result becomes unreliable at very low frequency $\omega \ll 1/\beta$ when the  spectral weight is relatively small. 

%

\begin{figure}[ht]
\includegraphics[clip,angle=0,width=8.3cm]{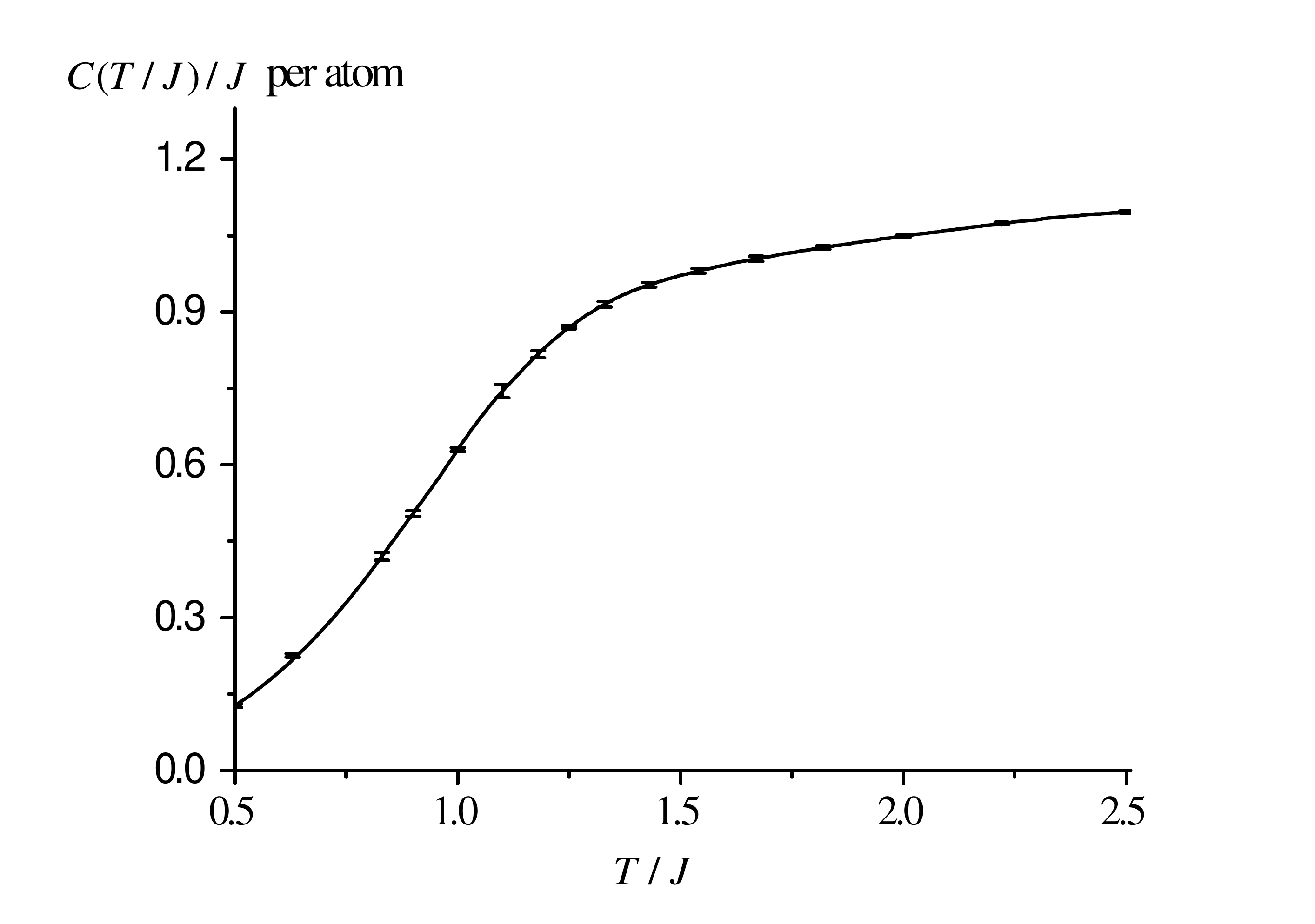}
\caption{\label{C-diag} Heat capacity $C(T/J)$ per atom as a function of temperature $T/J$.  For the homogeneous system, the Berezinskii\text{-}Kosterlitz\text{-}Thouless transition temperature is at $k_{\rm B} T_c \simeq 1.04J$. ~\cite{QCP2} Notice that below $T_c$, the heating of the system gets boosted due to the smallness of $C(T/J)$ .
}
\end{figure}

The heat capacity for a canonical-ensemble system has also been calculated and is shown in Fig. \ref{C-diag}. It is seen that the heat capacity becomes much smaller in the superfluid phase than in the normal phase, which may lead to more rapid heating.

\begin{figure}[ht]
\includegraphics[clip,angle=0,width=8.3cm]{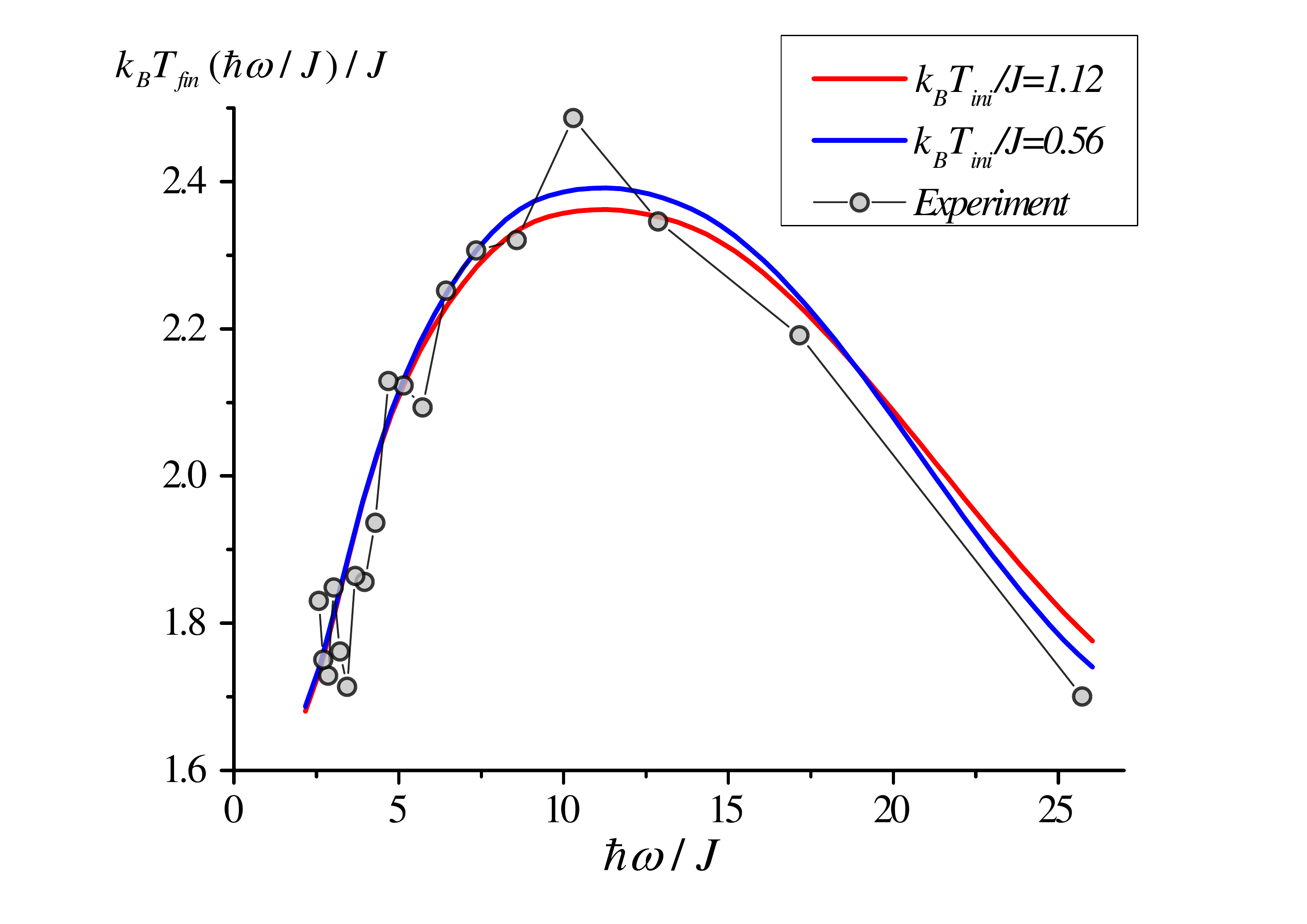}
\caption{\label{final-diag} Temperature response to the lattice-depth modulation that reveals the spectral function for the ultracold atomic system. The vertical axis represents final temperature. Filled circles connected by a black dashed line are final temperatures measured experimentally~\cite{bloch}. The solid lines (red and blue) are two predictions based on numerical calculations, with parameters $(k_{\rm B} T_{\rm ini}=0.56J, P=0.27J/{\rm sec})$ and  $(k_{\rm B} T_{\rm ini}=1.12J, P=0.45J/{\rm sec})$ respectively. In the calculations, we assume that the heating power $P$ is independent of temperature.
}
\end{figure}

To solve Eq.(\ref{eq:final}) self-consistently, the initial temperature $T_{\rm ini}$ and the heating power $P(T)$ are also required. However, both quantities were not addressed by the previous experiment nor is $P(T)$ computable by Monte Carlo simulations. Thus, we are forced to consider both quantities as fitting parameters. In Fig. \ref{final-diag}, we show two possible temperature responses to modulation obtained by solving Eq. (\ref{eq:final}), which both fit the experimental data well despite having rather different (but realistic) sets of $(T_{\rm ini}, P)$. Excellent agreement between the numerical and experimental results not only ensures that the analytical continuation procedure
(as routinely applied on the kinetic energy correlation function~\cite{Lode,Kun,Snir1,Snir2,Rancon,rose})
is reliable, but also validates various assumptions made in the first-principle calculation, 
such as quasi-static thermodynamics.
We also would like to point out that there are no fundamental difficulties in experiment to measure $T_{\rm ini}$ and $P(T)$, and thus an even more stringent test avoiding any fitting can be 
envisioned in the future.
%

%
%
%
%
\section{On the experimental observation of the massive Goldstone peak\label{sec:observation}}
%
%

Comparing the temperature response in Fig. \ref{final-diag} with the spectral function for a homogeneous system with $j/j_c=1.2$ (or $U=14J$) (see Fig. 2 of Ref. \onlinecite{Lode}), we find that while the steep onset of the spectral weight correlates remarkably well with the GM-mode energy, 
the resonance structure is lost in the experimental system. As mentioned previously, this may occur for two unrelated reasons: either because the MG-mode is overdamped, or the resonant signal is broadened by finite temperature and system inhomogeneity (tight confinement) effects~\cite{Lode}.

\begin{figure}[ht]
\includegraphics[clip,angle=0,width=8.3cm]{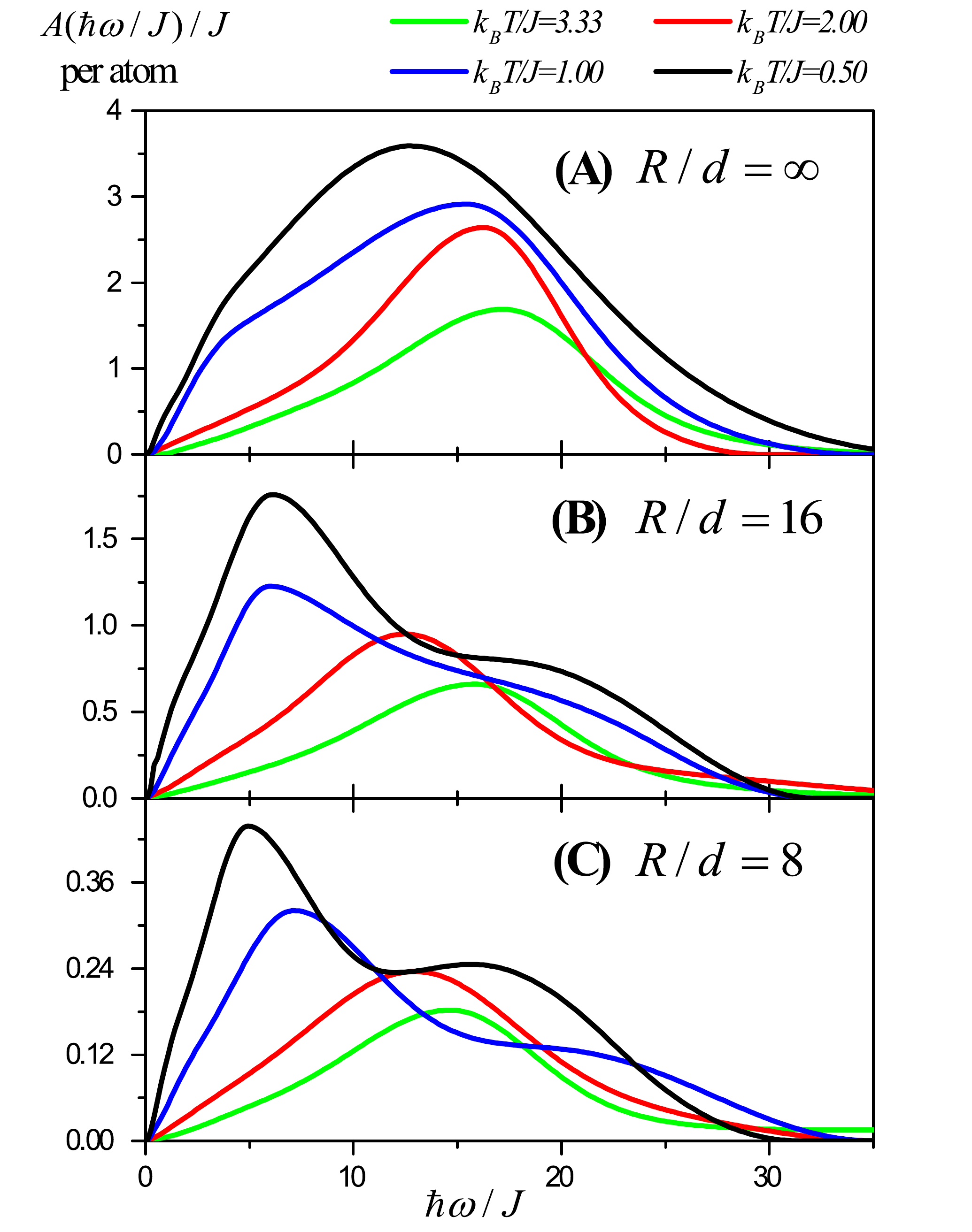}
\caption{\label{changeR-diag} Spectral functions for different modulation areas (a square of size $R \times R$ whose center coincides with the trap center) and temperatures for a system with 
with $N=800$ atoms and $j/j_c=1.05$. The  MG-resonance emerges at temperatures 
$T \sim J/k_{\rm B}$ when the modulation is limited to a mesoscopic area of linear 
size $R=16d$, where $d$ is the lattice spacing.
}
\end{figure}

Our simulations indicate that the second scenario is far more likely. Previous work on the homogeneous case established that a detuning  smaller than $j/j_c=1.05$ (or $U=16$) is required to clearly see the MG-resonance on top of the high-frequency continuum. Let us therefore take a system with  $j/j_c=1.05$, particle number $N=800$ and unity filling factor at the trap center 
and perform a numerical thought experiment: In order to reduce inhomogeneity effects, 
we limit the lattice-depth modulation to a mesoscopic area around the trap center, where the confining potential is nearly flat. For simplicity, we choose a square area with side length $R$.
In Fig.\ref{changeR-diag}, we show spectral functions for different values $R$. Resolving the resonance structure hiding in the inhomogeneous signal is dramatically improved 
when the modulation area is reduced. Though no resonance structure is seen when 
the entire system is modulated ($R/d=\infty$), it emerges when the modulation region is reduced to $R/d=16$ or $R/d=8$ at low enough temperature $T \sim T_c$ (where $T_c \simeq 0.45J/k_{\rm B}$ is the BKT temperature for a homogeneous  model with $j/j_c=1.05$). Converting spectral density to temperature response does not change this observation qualitatively, see  Fig.~\ref{changeR-diag}, even though the contrast for observing the resonance feature is diminishing.   
This thought experiment demonstrates that by taking care of response homogeneity, detuning from the QCP, and temperature,
the MG-peak can be seen in the kinetic energy spectral function using existing technology.

%
%
%

%
\begin{figure}[ht]
\includegraphics[clip,angle=0,width=8.3cm]{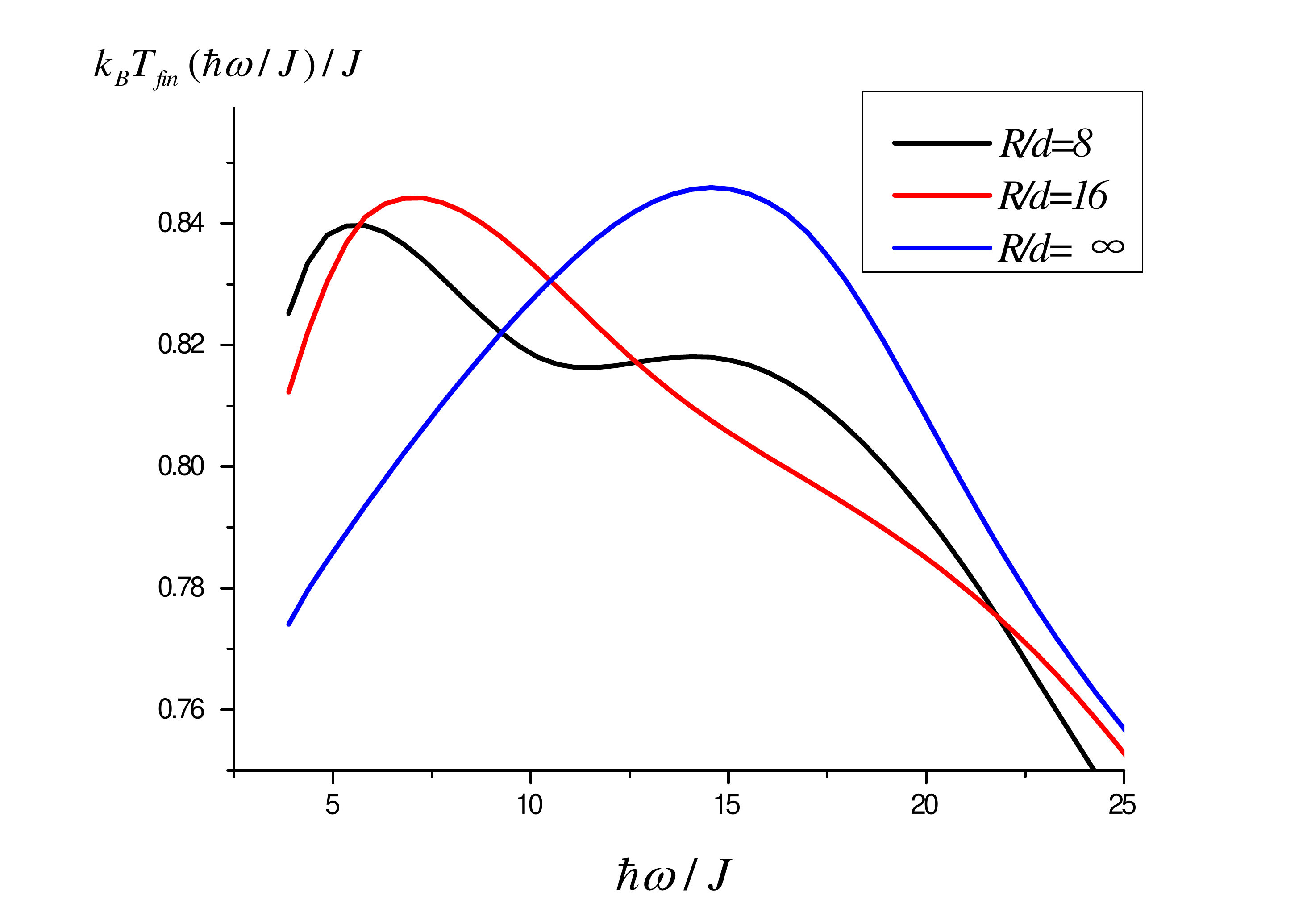}
\caption{\label{TchangeR} Temperature response for different modulation areas
for the same sets of model parameters as in Fig~\ref{changeR-diag}. We assume the system's heating power to be $P=0.2J/\text{\rm sec}$ and the initial temperature $k_{\text{\rm B}}T/J=0.5$. To optimize the contrast, the number of modulation cycles is set to be three and the sum of modulation and hold time is kept constant at $19$ milliseconds. The best modulation strength are found to be $\delta V_0/V_0=0.02, 0.03, 0.06$ for $R/d=8, 16, \infty$ respectively.}
\end{figure}
%
%

Combining the numerical results for the homogeneous and inhomogeneous model from Ref. \onlinecite{Lode, Kun} and from this work, we deduce that the following three conditions 
are to be met in order to reveal the MG-peak:

First, the modulation area should be restricted to the region with nearly constant 
chemical potential to ensure that the temperature response is measured for a 
homogeneous system. To achieve it, a straightforward approach would be to replace 
the harmonic confinement with the flat-bottom plus sharp walls potential. This approach, however,
may lead to problems with controlling system's density and entropy, and , thus, detuning 
from the QCP. An alternative approach is to restrict modulation to a mesoscopic area 
around the trap center, as shown is done in Fig. \ref{TchangeR}. One promising experimental implementation would be to apply a localized modulation of the scattering length~\cite{yoshiro}
using a laser beam induced Feshbach resonance~\cite{theis}. The technique has recently been shown in ultracold atom without optical lattice~\cite{yoshiro2010,cheng}. The size of the modulation area can be engineered by tuning the size of the laser beam (e.g. using a mask). Such a modulation would result in a time-dependent on-site interaction $U$ in the modulation area. As pointed out in Sec. \ref{sec:theory}, by subtracting a term proportional to $H_0$, the perturbation in 
potential energy can be replaced with the perturbation in kinetic energy, meaning that 
the MG-mode can be studied using the same temperature-response protocol as in the current experiment.

Second, the system has to be close enough to the QCP so that a Lorentz invariant action
provides an adequate description of physics. Our simulations indicate that a 
detuning $j/j_c \lesssim 1.05$ is sufficient to reveal the MG-resonance, 
while a smaller detuning $j/j_c \lesssim 1.02$ is required to recover the 
universal spectral function Eq. (\ref{eq:scaling}) including the critical 
pseudo-plateau at large frequencies~\cite{Kun}.

Third,  the system temperature has to be low enough. For a homogeneous system, simulations suggest that the resonance peak survives at temperatures as high as the BKT transition temperature $T_c$, but 
gradually goes away at $T>2T_c$~\cite{Lode}. Thus, having initial temperatures below $T_c$ is  recommended. For example, in the test system with $j/j_c=1.05$, $N=800$, and localized modulation size $R=8d$, which heats from $T_c$ up to $2T_c$ ( $T_c \simeq 0.45J/k_{\rm B}$), the 
resonance peak will remain visible in temperature response according to Fig. \ref{changeR-diag}(C).

To conclude, we would like to point out that the quantum critical dynamics in the MI and normal liquid phases is also of great interest. Numerical simulations indicate the presence of a universal resonance structure in the spectral function not only in the SF phase but also in phases with un-broken $U(1)$ symmetry, and at temperatures  $T \gg T_c$ (normal quantum critical liquid)~\cite{Kun}.
The existence of such universal resonances is unexpected within the current weak-coupling 
theory and their nature requires further study. Verification of this prediction 
from ultracold atom experiments would be crucial to solve this puzzle. 

%
%
%

%
%
%
\begin{center}
{\bf Acknowledgments}
\end{center}

KC thanks Yukawa Institute for Theoretical Physics at Kyoto University, where some of this work was done during the YITP-W-14-02 program on ``Higgs Modes in Condensed Matter and Quantum Gases''.  ME acknowledges support from the Harvard Quantum Optics Center. We also thank Immanuel Bloch, Andrey S. Mishchenko, Yuan Huang, Takeshi Fukuhara, Yoshiro Takahashi for valuable discussions. This work was supported in part by the National Science Foundation under grant No. PHY-1314735,  FP7/Marie-Curie Grant No. 321918 (``FDIAGMC"), FP7/ERC Starting Grant No. 306897 (``QUSIMGAS"), NNSFC Grant No. 11275185, CAS, NKBRSFC Grant No. 2011CB921300 and AFOSR/DoD MURI ``Advanced Quantum Materials: A New Frontier for Ultracold Atoms" program. We also thank the hospitality of the Aspen Center for Physics (NSF Grant No. 1066293).


\begin{thebibliography}{99}
%
%
%

\bibitem{book}
S. Sachdev, {\it Quantum Phase Transitions}, 2nd ed. (Cambridge University Press, Cambridge, 2011).

\bibitem{fisher}
M. P. A. Fisher, P. B. Weichman, G. Grinstein, and D. S. Fisher, Phys. Rev. B {\bf 40}, 546 (1989).

\bibitem{Goldstone}
 J. Goldstone, Nuovo Cim {\bf 19}, 154 (1961).

\bibitem{old_papers}
A. V. Chubukov, S. Sachdev, and J. Ye, Phys. Rev. B {\bf  49}, 11 919 (1994), S. Sachdev, Phys. Rev. B {\bf 59}, 14054 (1999), W. Zwerger, Phys. Rev.  Lett. {\bf 92}, 027203 (2004)

\bibitem{huber1}
S. D. Huber, E. Altman, H. P. B{\"u}chler, and G. Blatter,
Phys. Rev. B {\bf 75}, 085106 (2007).

\bibitem{huber2}
S. D. Huber, B. Theiler, E. Altman, and G. Blatter,
Phys. Rev. Lett. {\bf 100}, 050404 (2008).

\bibitem{pod11}
D. Podolsky, A. Auerbach, and D. P. Arovas, Phys. Rev. B {\bf 84}, 174522 (2011).


\bibitem{Machta}
E. Burovski, J. Machta, N.V. Prokof'ev,  and  B.V. Svistunov, Phys. Rev. B {\bf 74} 132502 (2006).

\bibitem{Vicari} M. Campostrini, M. Hasenbusch, A. Pelissetto, and E. Vicari, Phys. Rev. B {\bf 74}, 144506 (2006).


\bibitem{Lode}
L. Pollet and N. Prokof'ev, Phys. Rev. Lett. {\bf 109}, 010401 (2012).


\bibitem{Snir1}
S. Gazit, D. Podolsky, and A. Auerbach, Phys. Rev. Lett.
{\bf 110}, 140401 (2013),

\bibitem{Kun}
K. Chen, L. Liu, Y. Deng, L. Pollet, N. Prokof'ev, Phys. Rev. Lett. {\bf 110}, 170403 (2013)

\bibitem{Snir2}
S. Gazit, D. Podolsky, A. Auerbach, and D. P. Arovas, Phys. Rev. B {\bf 88}, 235108 (2013).

\bibitem{pod12}
D. Podolsky and S. Sachdev, Phys. Rev. B {\bf 86}, 054508 (2012).

\bibitem{Rancon}
A. Rancon and N. Dupuis, Physical Review B {\bf 89}, 180501 (2014).

\bibitem{rose}
 F. Rose, F. L\'{e}onard, and N. Dupuis,  Physical Review B {\bf 91}, 224501 (2015).
\bibitem{maxent}
M. Jarrell and J. Gubernatis, Phys. Rep. {\bf 269}, 133 (1996)

\bibitem{nikolay}
N. V. Prokof'ev and B. V. Svistunov, Jetp Lett. {\bf 97}, 649 (2013).

\bibitem{bloch}
M. Endres, T. Fukuhara, D. Pekker, M. Cheneau, P. Schau\ss, C. Gross, E. Demler, S. Kuhr, and I. Bloch,
Nature {\bf 487}, 454-458 (2012)

\bibitem{lode12}
L. Pollet, Rep. Prog. Phys. {\bf 75}, 094501 (2012).

\bibitem{spielman}
I. B. Spielman, W. D. Phillips, and J. V. Porto; Phys. Rev. Lett. 98 080404 (2007)

\bibitem{bloch_review}
 I. Bloch, J. Dalibard, and W. Zwerger, Rev. Mod. Phys. {\bf 80}, 885 (2008).

\bibitem{zoller}
D. Jaksch, C. Bruder, J. Cirac, C. Gardiner, and P. Zoller, Phys. Rev. Lett. {\bf 81}, 3108 (1998).

\bibitem{QCP1}
N. Elstner and H. Monien, Phys. Rev. B {\bf 59}, 12184 (1999).

\bibitem{QCP2}
B. Capogrosso-Sansone, S. G. S\"{o}yler, N. V. Prokof'ev, and
B. V. Svistunov, Phys. Rev. A {\bf 77}, 015602 (2008).

\bibitem{QCP3}
S. G. S\"{o}yler, M. Kiselev, N. V. Prokof'ev, and B. V.
Svistunov, Phys. Rev. Lett. {\bf 107}, 185301 (2011).

\bibitem{modulation}
T. St\"{o}ferle, H. Moritz, C. Schori, M. K\"{o}hl, and T. Esslinger, Phys. Rev. Lett. {\bf 92}, 130403 (2004).

\bibitem{Pollet2010}
L. Pollet, N. V. Prokof'ev, and B. V. Svistunov, Phys. Rev. Lett. {\bf 104}, 245705 (2010).

\bibitem{worm}
N. V. Prokof'ev, B. V. Svistunov, and I. S. Tupitsyn,Phys.
Lett. A {\bf 238}, 253 (1998), N. V. Prokof'ev, B. V. Svistunov, and I. S. Tupitsyn, Sov. Phys. JETP {\bf 87}, 310 (1998), L. Pollet, K. Van Houcke, and S. Rombouts, J. Comput. Phys. {\bf 225}, 2249 (2007)

\bibitem{yoshiro}
Yoshiro Takahashi, private communication

\bibitem{theis}
M. Theis, G. Thalhammer, K. Winkler, M. Hellwig, G. Ruff, R. Grimm, and J. H. Denschlag, Phys. Rev. Lett. {\bf 93}, 123001 (2004).

\bibitem{yoshiro2010}
R. Yamazaki, S. Taie, S. Sugawa, Y. Takahashi,
 Phys. Rev. Lett. {\bf 105}, 050405 (2010)
 
\bibitem{cheng}
L. W. Clark, L. C. Ha, C. Y.  Xu, C. Chin, arXiv. 1506.01766 (2015)


%
%
\end{thebibliography}
\end{document}